\documentclass[10pt]{article}
\usepackage{graphicx}
\usepackage{amssymb}

\def\Title#1{\begin{center} {\Large #1 } \end{center}}
\def\Author#1{\begin{center}{ \sc #1} \end{center}}
\def\Address#1{\begin{center}{ \it #1} \end{center}}

\newenvironment{Abstract}{\begin{quotation} \begin{center} 
             \large ABSTRACT \end{center}\bigskip 
      \begin{center}\begin{large}}{\end{large}\end{center} \end{quotation}}

\newenvironment{Presented}{\begin{quotation} \begin{center} 
             PRESENTED AT\end{center}\bigskip 
      \begin{center}\begin{large}}{\end{large}\end{center} \end{quotation}}

\def\Acknowledgements{\bigskip  \bigskip \begin{center} \begin{large}
             \bf ACKNOWLEDGEMENTS \end{large}\end{center}}

\def\beq{\begin{equation}}
\def\eeq#1{\label{#1}\end{equation}}
\def\eeqn{\end{equation}}

\def\beqa{\begin{eqnarray}}
\def\eeqa#1{\label{#1}\end{eqnarray}}
\def\eeqan{\end{eqnarray}}

\let\bar=\overbar

\def\Dslash{\not{\hbox{\kern-4pt $D$}}}
\def\dslash{\not{\hbox{\kern-2pt $\del$}}}

\def\msb{{\bar{\ssstyle M \kern -1pt S}}}

\usepackage{color}

\begin{document}

\large
\begin{titlepage}

\vfill
\Title{  THE WRONG SIGN LIMIT IN THE 2HDM }
\vfill

%
\Author{P.M.~Ferreira}
\Address{Instituto Superior de Engenharia de Lisboa - ISEL,
	1959-007 Lisboa, Portugal}
\Address{Centro de F\'{\i}sica Te\'{o}rica e Computacional,
    Faculdade de Ci\^{e}ncias,
    Universidade de Lisboa,
    Av.\ Prof.\ Gama Pinto 2,
    1649-003 Lisboa, Portugal}
\Author{Renato Guedes}
\Address{ Centro de F\'{\i}sica Te\'{o}rica e Computacional, Faculdade de Ci\^{e}ncias, Universidade de Lisboa,
        Av.\ Prof.\ Gama Pinto 2, 1649-003 Lisboa, Portugal}
\Author{John F.~Gunion}
\Address{  Davis Institute for High Energy Physics,
    University of California,
    Davis, California 95616, USA}
\Author{Howard E.~Haber}
\Address{ Santa Cruz Institute for Particle Physics,
    University of California and \\
       Ernest Orlando Lawrence Berkeley National Laboratory,
University of California, Berkeley, California 94720, USA}
\Author{Marco O.P.~Sampaio}
\Address{ Departamento deF\'{\i}sica da Universidade de Aveiro and I3N
        Campus de Santiago, 3810-183 Aveiro, Portugal}
\Author{Rui Santos}
\Address{Instituto Superior de Engenharia de Lisboa - ISEL,
	1959-007 Lisboa, Portugal}
\Address{Centro de F\'{\i}sica Te\'{o}rica e Computacional,
    Faculdade de Ci\^{e}ncias,
    Universidade de Lisboa,
    Av.\ Prof.\ Gama Pinto 2,
    1649-003 Lisboa, Portugal}
\vfill

\begin{Abstract}
\begin{center}
A sign change in the Higgs couplings to fermions and massive gauge bosons is still allowed in the framework of 
two-Higgs doublet models (2HDM). In this work we discuss the possible sign changes in the Higgs couplings to fermions
and gauge bosons, while reviewing
the status of the  8-parameter CP-conserving 2HDM after the Large Hadron Collider 8 TeV run.
\end{center}
\end{Abstract}
\vfill

\begin{Presented}
The Second Annual Conference\\
 on Large Hadron Collider Physics \\
Columbia University, New York, U.S.A \\ 
June 2-7, 2014
\end{Presented}
\vfill
\end{titlepage}
\def\thefootnote{\fnsymbol{footnote}}
\setcounter{footnote}{0}
%

\normalsize 


\section{Introduction}
The ATLAS~\cite{ATLASHiggs} and CMS~\cite{CMSHiggs} collaborations at the Large Hadron Collider (LHC)
have finally confirmed the existence of a Higgs boson. The measurements of the couplings to the fermions
and gauge bosons have shown no major deviations from the Standard Model (SM) predictions. Also, because
no other scalars were found, the parameter space of most extensions of the SM is becoming heavily constrained.  
Such is the case of the model where an extra doublet is added to the SM content, known as the
 two-Higgs doublet model (2HDM).

We will show that the parameter space of the 2HDM is indeed constrained to be close to the SM predictions
as to reproduce the experimental results obtained at the LHC at the end of the 8 TeV run. There are however
some interesting regions of parameter space where the Higgs couplings to fermions or gauge bosons change
sign relative to the SM Higgs couplings. Whenever this sign change is measurable, given enough experimental
precision, we call it wrong sign scenario~\cite{Ferreira:2014naa}. This scenario was first studied in~\cite{Ginzburg:2001ss}
and more recently in~\cite{Dumont:2014wha, Fontes:2014tga}. 

\section{The model}

The 2HDM is built by simply adding a complex scalar doublet to the SM field content. In its most general form, the
2HDM Yukawa Lagrangian gives rise to tree-level Higgs-mediated flavour-changing neutral 
currents (FCNCs), in disagreement with experimental data. There is however a simple way to avoid these 
tree-level FCNCs which is to impose a $Z_2$ symmetry on the two
scalar doublets, $\Phi_1 \rightarrow \Phi_1$, $\Phi_2 \rightarrow - \Phi_2$. By imposing a corresponding 
discrete symmetry on the fermion fields one is able to construct four independent Yukawa models types I, II, Flipped (F) (or Y)
and Lepton Specific (LS) (or X). In type I only $\Phi_2$ couples to all fermions; in type II $\Phi_2$ couples to up-type quarks and $\Phi_1$ couples to 
down-type quarks and leptons; in type F (Y) $\Phi_2$ couples to up-type quarks and to leptons and $\Phi_1$ couples to down-type quarks; finally
in type LS (X) $\Phi_2$ couples to all quarks and $\Phi_1$ couples to leptons. See~\cite{hhg} for a comprehensive review on the 2HDM.

We will work with a softly broken $Z_2$ symmetric scalar potential that can be written as
\begin{eqnarray}
V(\Phi_1,\Phi_2) =& m^2_{11} \Phi^{\dagger}_1\Phi_1+m^2_{22}
\Phi^{\dagger}_2\Phi_2 - (m^2_{12} \Phi^{\dagger}_1\Phi_2+{\mathrm{h.c.}
}) +\frac{1}{2} \lambda_1 (\Phi^{\dagger}_1\Phi_1)^2 +\frac{1}{2}
\lambda_2 (\Phi^{\dagger}_2\Phi_2)^2\nonumber \\ 
&+ \lambda_3
(\Phi^{\dagger}_1\Phi_1)(\Phi^{\dagger}_2\Phi_2) + \lambda_4
(\Phi^{\dagger}_1\Phi_2)(\Phi^{\dagger}_2\Phi_1) + \frac{1}{2}
\lambda_5[(\Phi^{\dagger}_1\Phi_2)^2+{\mathrm{h.c.}}] ~, \label{higgspot}
\end{eqnarray}
where $\Phi_i$, $i=1,2$ are complex SU(2) doublets. We focus on a specific realisation
of the 2HDM, the usual 8-parameter CP-conserving potential where both the potential parameters and the VEVs are real.
We choose as free parameters the four masses $m_h$, $m_H$, $m_A$ and $m_{H^\pm}$, the rotation
angle in the CP-even sector, $\alpha$, the ratio of the vacuum expectation
values,  $\tan\beta=v_2/v_1$, and the soft breaking parameter $m_{12}^2$.
We choose as a convention for the angles,  without loss of generality, $0\leq\beta\leq \pi/2$ and $- \pi/2 \leq  \alpha \leq \pi/2$.

The addition of an extra doublet could lead to the breaking of $U(1)_{em}$
and the consequent non-conservation of electric charge. However, it was shown that  
the existence of a tree-level scalar potential minimum that breaks the electroweak symmetry but 
preserves both the electric charge and CP symmetries, ensures that no additional tree-level potential minimum
that spontaneously breaks the electric charge and/or CP symmetry
can exist~\cite{vacstab}. Finally, we force the CP-conserving minimum to be the global one~\cite{Barroso:2013awa}.

\section{Results and discussion}

In this section we present the parameter space of the 2HDM that is still allowed after the 8 TeV run
for the different Yukawa types. A scan in this space was performed with ScannerS~\cite{Coimbra:2013qq} 
with the parameters in the following ranges:
$m_h = 125.9~{\rm GeV}$, 
$m_h + 5~{\rm GeV} <m_H,\, m_A < 1~{\rm TeV}$, $100~{\rm GeV} < m_{H^\pm} < 1~{\rm TeV}$, 
$1 < \tan \beta < 50$, $|\alpha| < \pi/2$
and  $- (900~{\rm GeV})^2<m_{12}^2 <(900~{\rm GeV})^2$. 
ScannerS is interfaced with SusHi~\cite{Harlander:2012pb} and HDECAY~\cite{Djouadi:1997yw, Harlander:2013qxa}
for Higgs production and decays, cross-checked with HIGLU~\cite{Spira:1995mt} and 2HDMC~\cite{Eriksson:2009ws}.
The remaining Higgs production cross sections were taken from~\cite{LHCHiggs}. Collider data was taken into
account with HiggsBounds~\cite{Bechtle:2013wla} and HiggsSignals~\cite{Bechtle:2013xfa}. The
remaining constraints (see~\cite{Barroso:2013zxa}), theoretical, electroweak precision and B-physics constraints are coded in ScannerS.

We define
\begin{equation}
\kappa_i^2=\frac{\Gamma^{\scriptscriptstyle {\rm 2HDM}}  (h \to i)}{\Gamma^{\scriptscriptstyle {\rm SM}} (h \to i)}
\end{equation}
which at tree-level is just the ratio of the couplings $\kappa_i=g_{i}^{\scriptscriptstyle {\rm 2HDM}}  /g_{i}^{\scriptscriptstyle {\rm SM}} $.
We will also use the standard definition of signal strength
\begin{equation}
\mu_f \, = \, \frac{\sigma \, {\rm BR} (h \to
  f)}{\sigma^{\scriptscriptstyle {\rm SM}} \, {\rm BR^{\scriptscriptstyle{\rm SM}}} (h \to f)}
\label{eg-rg}
\end{equation}
where $\sigma$ is the Higgs production cross section and ${\rm BR} (h \to f)$ is
the branching ratio of the decay into some given final state $f$;  $\sigma^{\scriptscriptstyle {\rm {SM}}}$
and ${\rm BR^{\scriptscriptstyle {\rm SM}}}(h \to f)$ are the expected values for the same quantities in the SM. 
We will not separate different LHC initial state production mechanisms and instead sum over all production mechanisms in computing the cross section.

\begin{figure}[h!]
\centering
\includegraphics[width=0.37\linewidth]{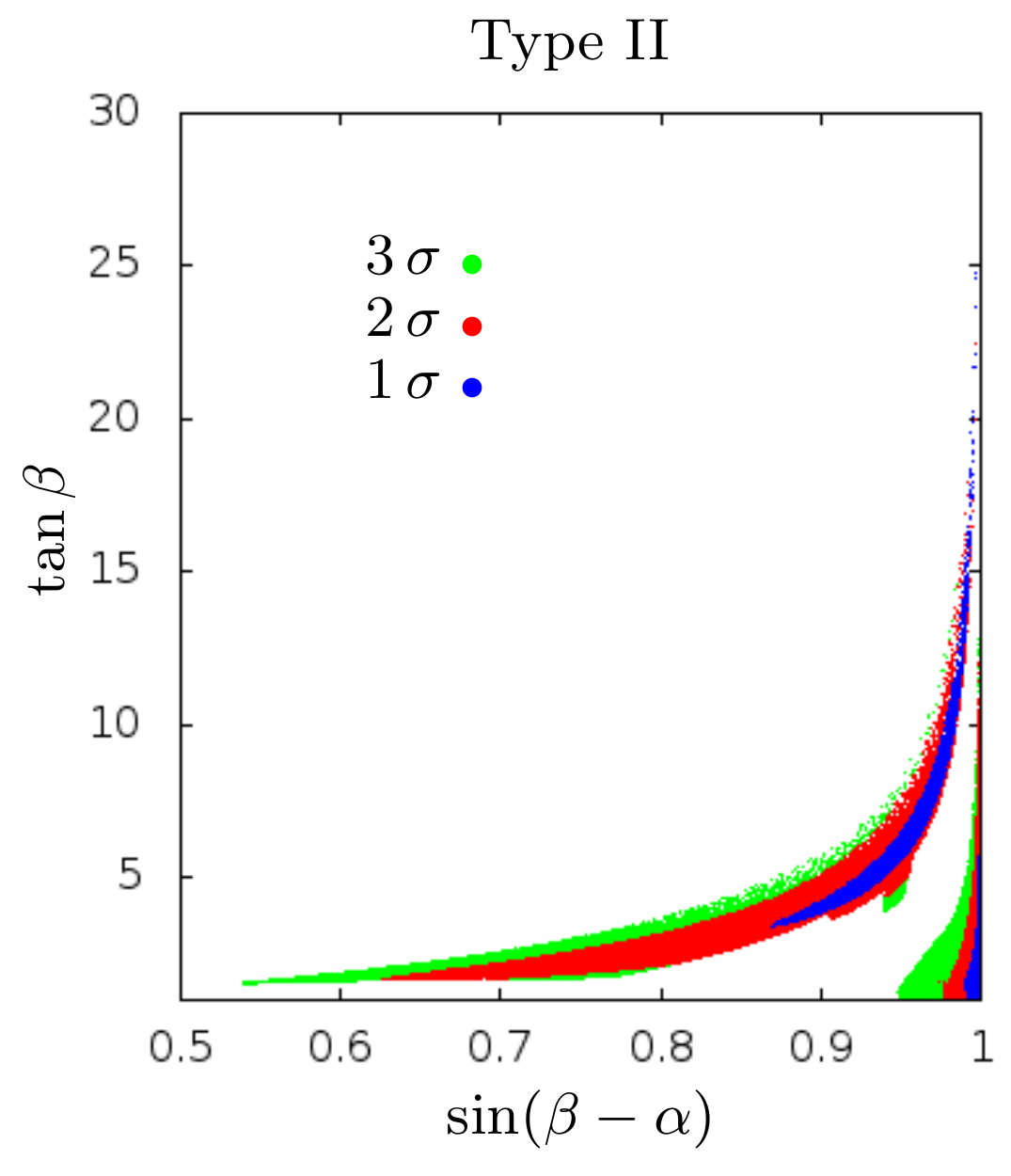}\hspace{0.09\linewidth}\includegraphics[width=0.373\linewidth]{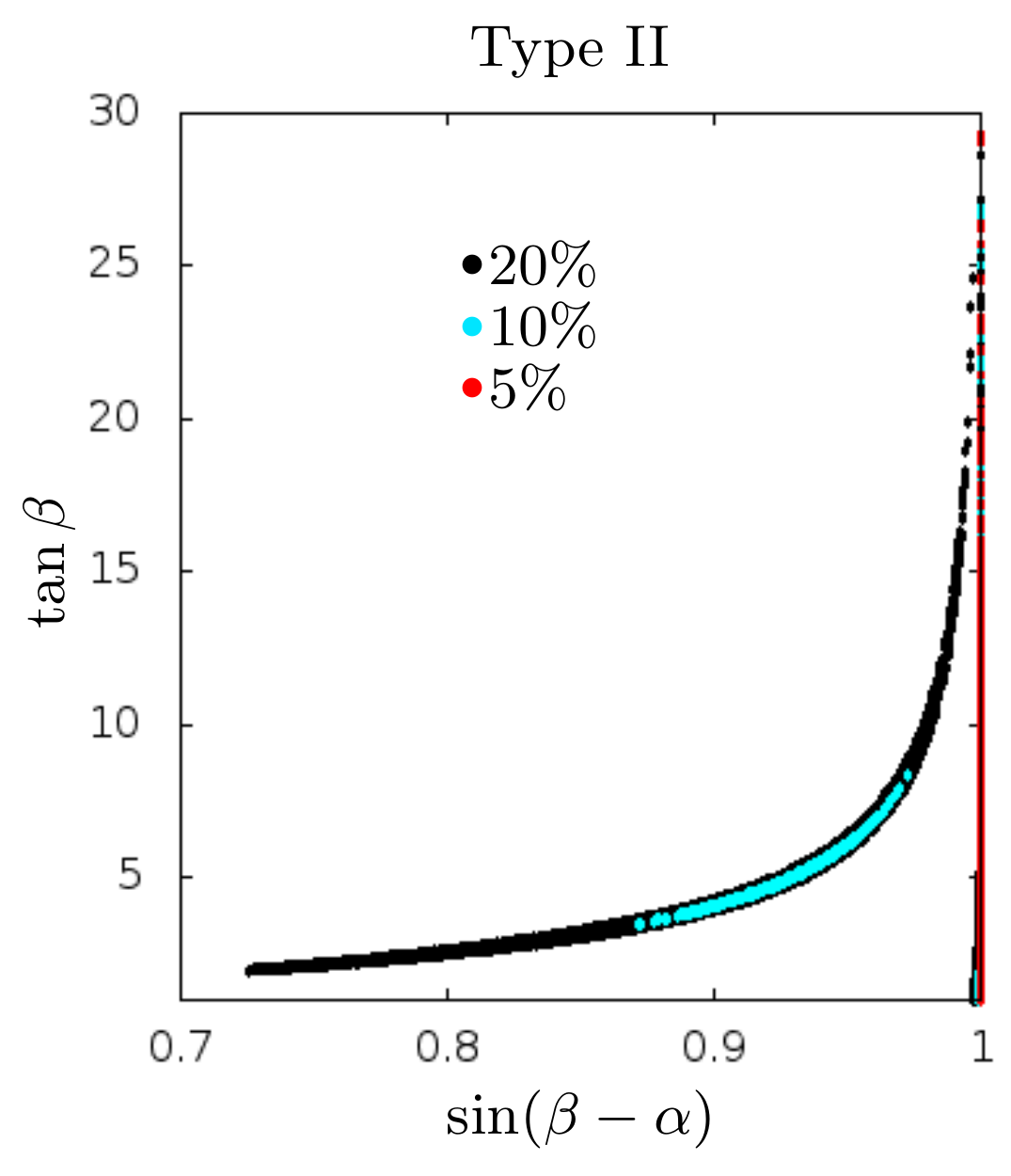}
\caption{$\tan \beta$ as a function of $\kappa_V = \sin (\beta -\alpha)$ for the type II model. Left panel: allowed parameter space taking into account
all theoretical constraints and all experimental data including all data from the LHC's 8 TeV run analysed so far at $1$ (blue), $2$ (red) and $3 \sigma$ (green); right panel: same as left panel except
for the LHC constraints, where instead we ask that all measured rates ($\mu_{VV}$, $\mu_{\gamma \gamma}$ and $\mu_{\tau \tau}$) are within $5$ (red), $10$ (blue) and $20 \%$ (black) of the SM predictions. }
\label{fig:T1}
\end{figure}

In the left panel of figure~\ref{fig:T1} we present the allowed parameter space of the type II CP-conserving 2HDM after the 8 TeV run at $1$, $2$ and $3\sigma$ 
with all experimental and theoretical constraints taken into account. The most striking feature in the plot is that there are two distinct regions that come together
for $\tan \beta$ close to $\approx 17$.  The region on the the right corresponds to the the SM-like or alignment limit, that is, $\sin (\beta - \alpha) =1$.
In this limit, the lightest Higgs couplings to gauge bosons and to fermions are the SM ones. 
The region on the left is centred around the line $\sin (\beta + \alpha) =1$. 
In type II and with our conventions, it corresponds to the limit where the Higgs coupling to down-type
quarks changes sign relative to the SM while couplings to up-type quarks and massive gauge bosons remain the same.
In fact, when $\sin (\beta + \alpha) =1$, one obtains $\kappa_D = -1$ in type II, $\kappa_V \, \kappa_D <0$, which
was called the wrong sign limit in~\cite{Ferreira:2014naa} (a limit imposed at tree-level). 
The most important point to note here is that the two limits can be distinguished experimentally. In fact,
the alignment  limit is already squeezed into values of $\sin (\beta - \alpha) $ extremely close to $1$ even at $3\sigma$. On the contrary, the wrong sign limit spans
a much larger region of $\sin (\beta - \alpha) $, which at $3\sigma$ goes from about $0.52$ to $1$. The reason for the two regions meeting at large $\tan \beta$
is that because $\sin (\beta - \alpha) - \sin (\beta + \alpha) =(\tan^2 \beta -1)/(\tan^2 \beta +1) $,  $\sin (\beta - \alpha) \approx \sin (\beta + \alpha) $ for large $\tan \beta$. The shape of these lines is primarily determined by the LHC constraints on $\mu_{VV}$.
 
In the right panel we present $\tan \beta$ as a function of $\kappa_V = \sin (\beta -\alpha)$ with the same theoretical and experimental constraints
 except that the LHC present constraints  on the measured Higgs rates are replaced by asking that all measured rates for the final states 
$f=WW$, $ZZ$, $\gamma \gamma$ and $\tau^+ \tau^ -$, are within $5$ (red), $10$ (blue) and $20 \%$ (black) of the SM predictions.
Both the alignment limit and the wrong sign limit are still present. As expected, the two regions shrink and the alignment limit is now almost a straight line for $\sin (\beta -\alpha) =1$.
 However, two interesting features emerge. First, a measurement of the rates at $5\%$ will exclude the wrong sign scenario as discussed in~\cite{Ferreira:2014naa}. 
 Second, large $\tan \beta$ values are excluded for the wrong sign scenario at $10 \%$ in type II. This is due to the bound imposed on $\mu_{\tau^+ \tau^-}$ and it is 
 ultimately a consequence of the large values of the $(gg+bb) \to h$ production cross section in the wrong sign limit as compared to the SM (there is no significant change in the 
$h \to \tau^+ \tau^ -$ branching ratios).
 
\begin{figure}[h!]
\centering
\includegraphics[height=0.37\linewidth]{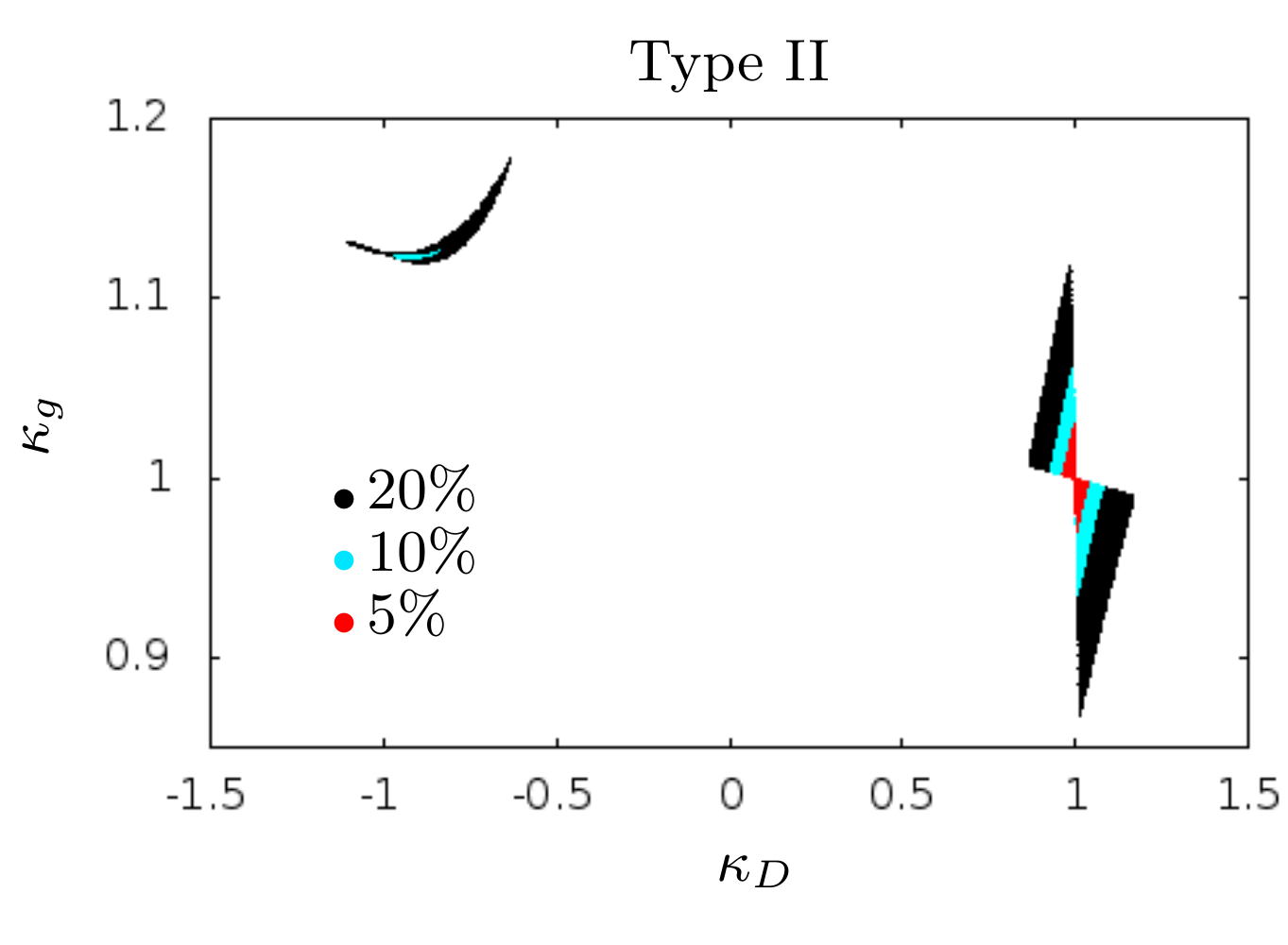}\hspace{0.05\linewidth}\includegraphics[height=0.395\linewidth]{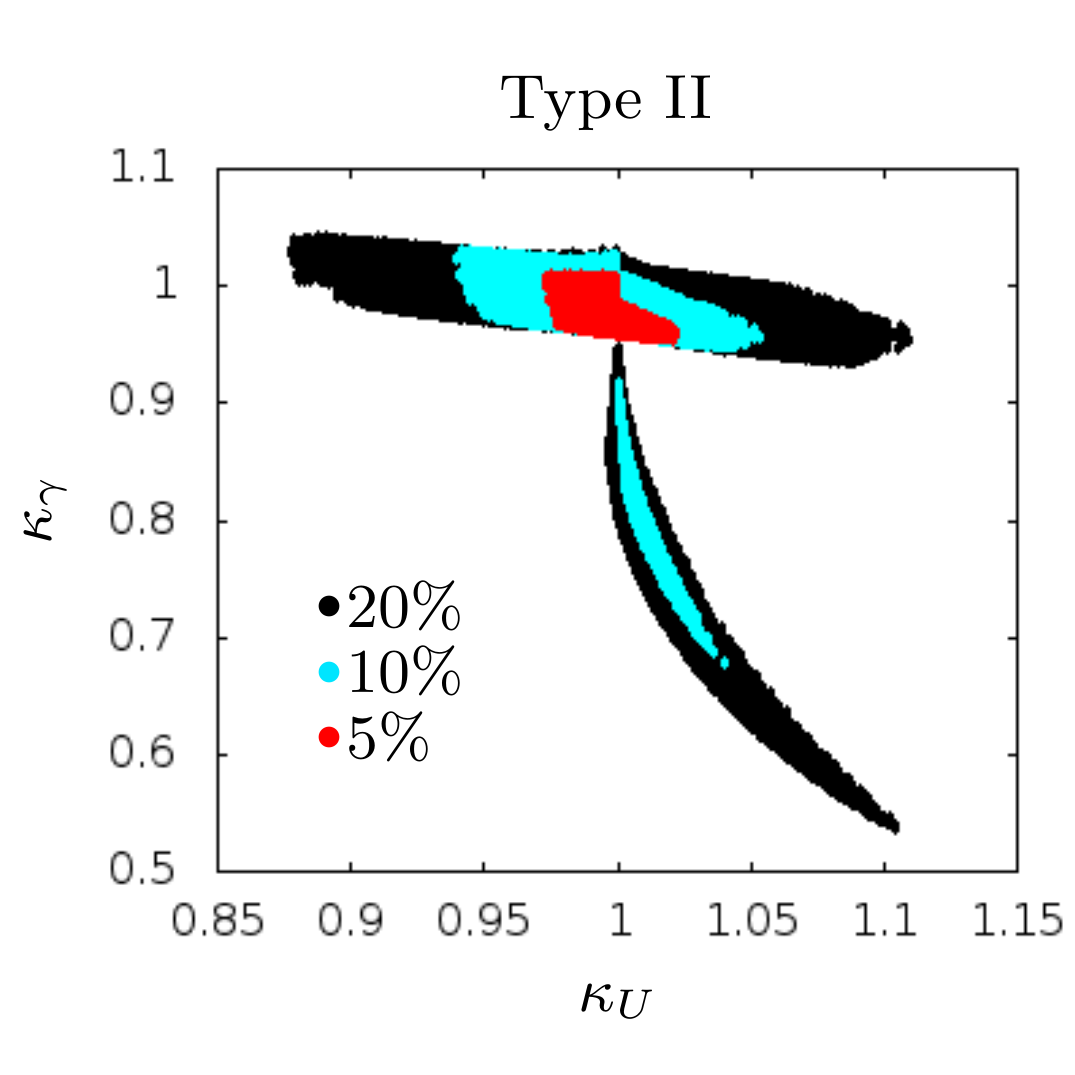}
\caption{Left panel: $\kappa_g$ as a function of $\kappa_D$. Right panel: $\kappa_\gamma$ as a function of $\kappa_U$ where the upper region
corresponds to the alignment limit while the lower one corresponds to the wrong sign limit.}
\label{fig:T2}
\end{figure}

When $\kappa_V = \sin (\beta - \alpha) =1$, $\kappa_U=\kappa_D=\kappa_L =1$, that is, the lightest Higgs couplings
to up-type quarks ($U$),  down-type quarks ($D$) and leptons ($L$), are the SM ones. The particular wrong sign scenario we are discussing, 
which occurs for $\tan \beta > 1$, is defined as $\kappa_D \, \kappa_V <0$ and $\kappa_D \, \kappa_U <0$. 

In the wrong sign limit, the Higgs production cross section via gluon fusion is enhanced due to the sign change in $\kappa_D$. 
The other modes that are different in the two limits (at LO) are the ones involving $\kappa_V$,
that is  VBF and associated $Vh$ production. As previously discussed, this difference vanishes for large $\tan \beta$.
Concerning the Higgs decays, there is no difference between the two scenarios in the decay to fermions. Again,
because $\sin (\beta - \alpha) \approx \sin (\beta + \alpha)$ for $\tan \beta \gg 1$, taking $\tan \beta = 8$ the ratio
of the wrong sign $\Gamma (h \to WW \, (ZZ))$ decay width to the respective SM width is 0.94, which corresponds
to a small effect in $\mu_{VV} $. A difference could however appear for small values of $\tan \beta$.
Finally, the decays $h \to \gamma \gamma$ and $h \to gg$ could present a meaningful difference between
the two limits due to the interference between the different loop contributions.
 
In figure~\ref{fig:T2} we present $\kappa_g$ as a function of $\kappa_D$ (left) and  $\kappa_\gamma$ as a function of
$\kappa_U$ in type II with all rates within $20\%$ (blue), $10\%$ (green) and $5\%$ (red) of the SM values.
As a change in sign in $\kappa_D$ leads to $\kappa_g \approx 1.13$, it is expected that a precise measurement of 
$\kappa_g$ could exclude in the future the wrong sign scenario ($\kappa_g$ is extracted indirectly since 
there is no direct measurement of $h \to gg$). Regarding $\kappa_\gamma$, shown on the right plot,
one can see two distinct regions: the upper region is the alignment limit and the lower region is the wrong sign limit
where  $\kappa_D=-1$. The change of the $\kappa_D$ sign in the SM amplitude 
amounts to only a $1\%$ difference in $\Gamma (h \to \gamma \gamma)$ relative
to the respective SM width. Therefore the larger difference seen in the left panel of figure \ref{fig:T2} has to come
from the charged Higgs contribution. As was shown in~\cite{Ferreira:2014naa} this contribution to $\Gamma (h \to \gamma \gamma)$
in the $\kappa_D <0$ case is approximately constant and always sufficiently significant as to eventually be observable at the LHC. One should note
however that the constraints coming from tree-level unitarity imply that the result is only perturbatively reliable for $m_{H^\pm} \lesssim 650$ GeV .

In order to understand if these differences can indeed be measured at the LHC we refer to table~1-20 of Ref.~\cite{Dawson:2013bba}.
It is shown that the expected errors for $\kappa_g$ based on fittings are 
$6$--$8\%$ for $L=300$ fb$^{-1}$ and $3$--$5\%$ for $L=3000$ fb$^{-1}$ (for 14 TeV).
The predicted
accuracy for $\kappa_\gamma$ is $5$--$7\%$ for an integrated luminosity of $L=300$ fb$^{-1}$ and $2$--$5\%$ for $L=3000$ fb$^{-1}$.
In view of the plots in figure~\ref{fig:T2} it is clear that there are good chances of probing the wrong-sign scenario in the 14 TeV LHC run. With the predicted accuracy
for the International Linear Collider~\cite{Ono:2012ah, Asner:2013psa} the wrong sign limit could not only be probed by 
a measurement of $\kappa_g$  and $\kappa_\gamma$ but also in the process $e^+ e^- \to Z h ( \to b \bar{b}) $. Finally one should 
note that a thorough study of this scenario has to take into account the 2HDM electroweak corrections, some
of which are already available~\cite{LopezVal:2009qy, Kanemura:2014dja}.

\begin{figure}[h!]
\centering
\includegraphics[width=0.37\linewidth]{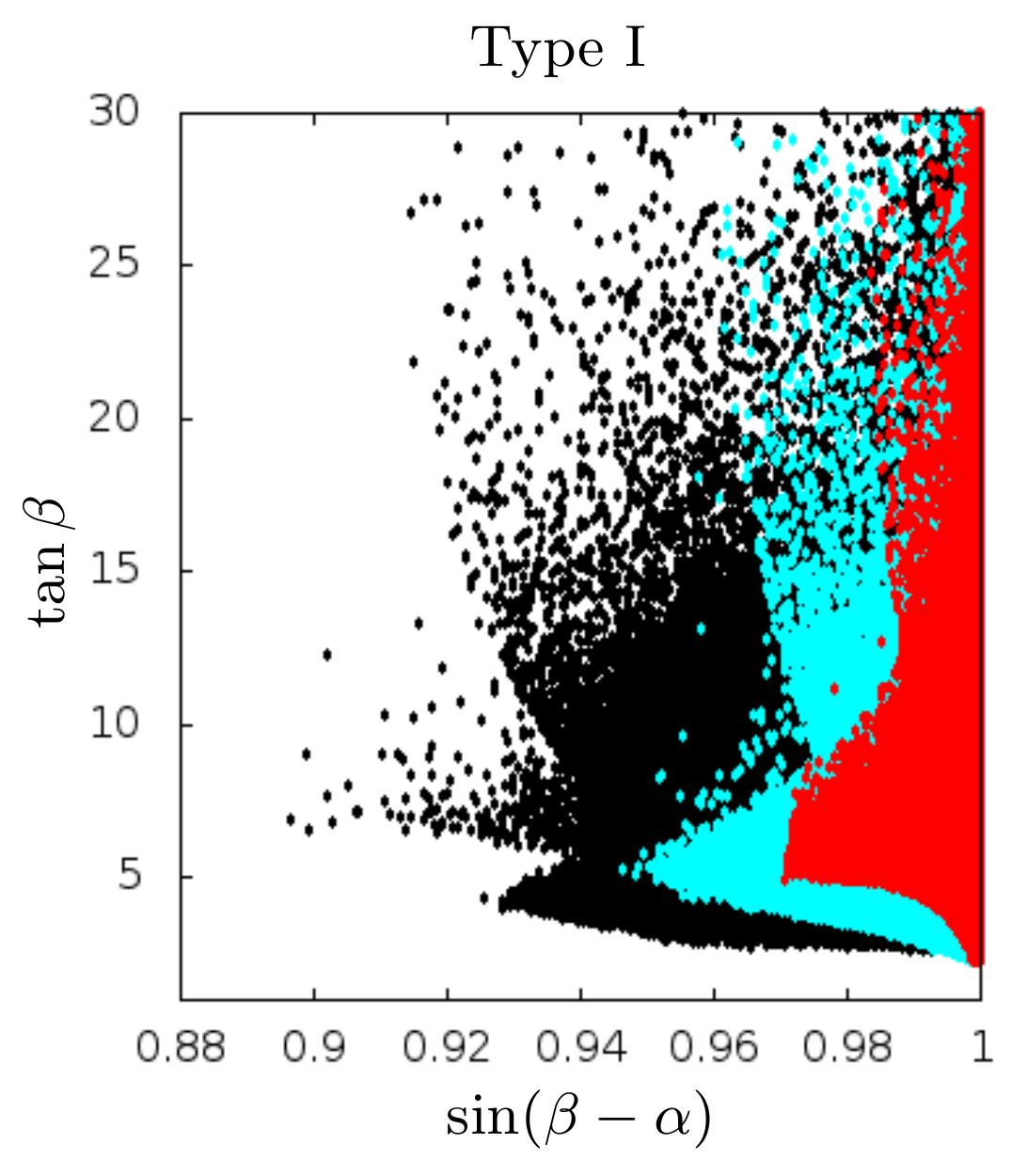}\hspace{0.12\linewidth}\includegraphics[width=0.37\linewidth]{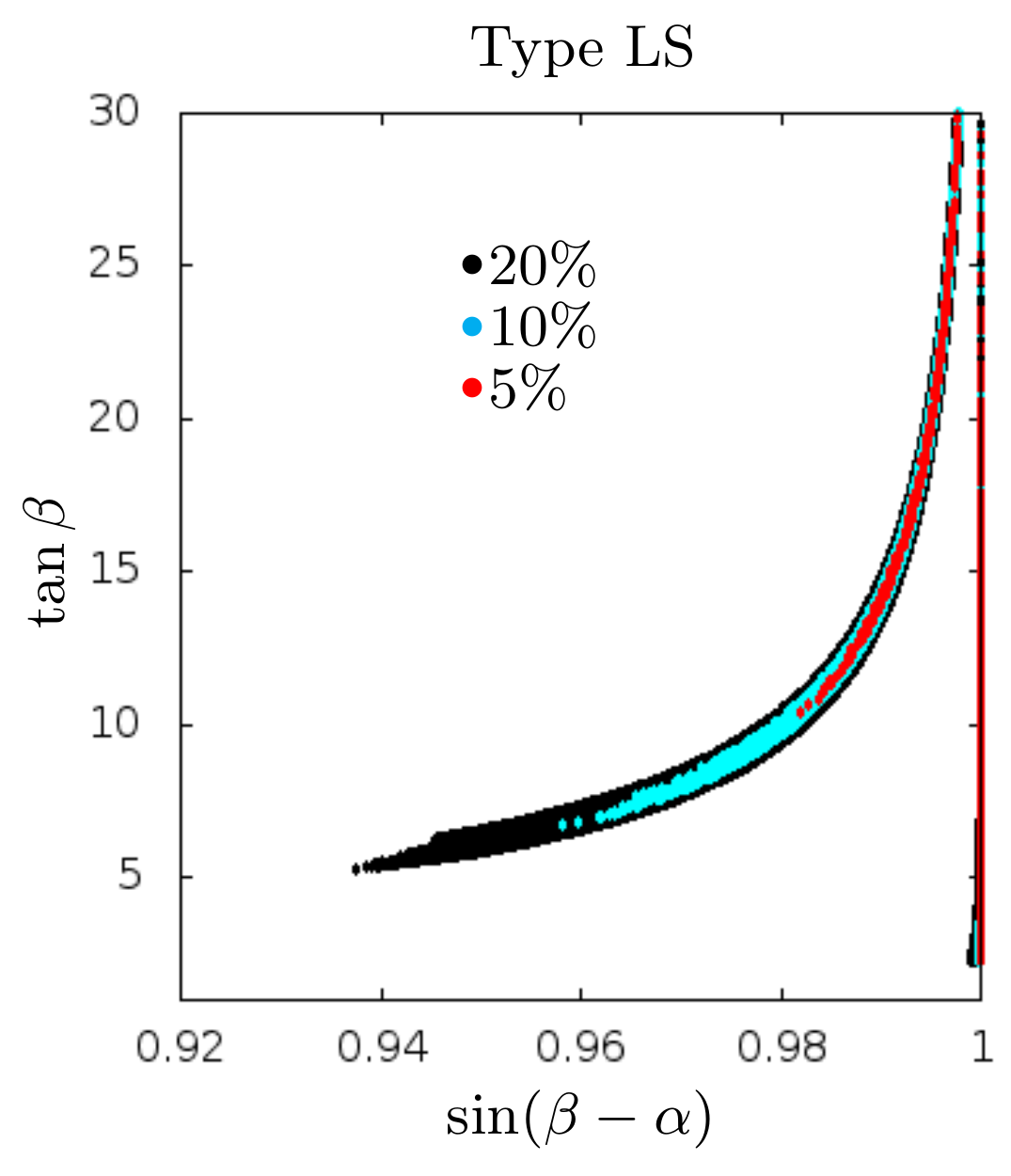}
\caption{$\tan \beta$ as a function of $\kappa_V = \sin (\beta -\alpha)$ for type I (left) and type LS (right) for the allowed parameter space taking into account
all theoretical constraints and all experimental data.}
\label{fig:T3}
\end{figure}

Finally, in figure~\ref{fig:T3}  we present $\tan \beta$ as a function of $\kappa_V = \sin (\beta -\alpha)$ for type I (left) and type LS (right) for the allowed parameter space taking into account
all theoretical constraints and all experimental data with all measured rates within $5$ (red), $10$ (blue) and $20 \%$ (black) of the SM predictions. 
For the type I model we see that the limit on $\sin (\beta - \alpha)$ improves due to the increasing precision on
the measurement of $\mu_{VV}$. The bound is almost independent of $\tan \beta$ because the Yukawa couplings are all rescaled by the same factor. Furthermore, for type I and for $\tan \beta \gg 1$,
we have $\kappa_V = \kappa_F = 1$. The deviation from a constant behaviour in $\tan \beta$ for type I is due to the remaining theoretical and experimental constraints.
In type LS, 
the $\tan \beta$ dependence comes both from the measurement of $\mu_{VV}$ and $\mu_{\tau \tau}$~\cite{Arhrib:2011wc}. Although there
is no wrong sign limit for type I and type LS, there is the \textit{symmetric limit} as described in~\cite{Ferreira:2014dya}.
This limit can be defined for type I 2HDM, where in the limit $\sin(\beta+\alpha)=1$, none of the Higgs couplings to the remaining SM particles changes sign relative to the SM one. Nevertheless, the 
shift $\alpha \to - \alpha$ still changes the value of $\kappa_V$, which is given by
\begin{equation}
\kappa_V=\frac{\tan^2 \beta -1}{\tan^2 \beta +1} \, .
\end{equation}
For type LS only the coupling to leptons changes sign and this change plays no role 
in the analysis.
Let us recall that $\sin (\beta - \alpha) \approx \sin (\beta + \alpha)$ only for large $\tan \beta$.
Therefore, the two curves on the right plot are close to the lines $\sin (\beta - \alpha)=1$ (right) and $\sin (\beta + \alpha)=1$ (left). Clearly the two curves can be distinguished, given enough precision,
except for high values of $\tan \beta$. In conclusion, for (some) low values of $\tan \beta$ the \textit{symmetric} limit can be probed at the next LHC run with very high luminosity, at least for type LS. The situation
is not as clear for type I because the two regions are superimposed until $\sin (\beta -\alpha)$ is almost 1.

\Acknowledgements
PMF, RG and RS are supported by FCT
under contracts PTDC/FIS
/117951/2010 and PEst-OE/FIS/UI0618
/2011.
RG is also supported by a FCT Grant SFRH/BPD/47348/2008.
MS is supported by a FCT Grant SFRH/BPD/-69971/2010.
JFG is supported in part by U.S. DOE grant DE-SC-000999.
HEH is support in part by U.S. DOE grant DE-FG02-04ER41286.


\end{document}